\documentclass[prd,aps,preprint,tightenlines,showpacs,nofootinbib,superscriptaddress]{revtex4-1}
\usepackage{mathrsfs}
\usepackage{amsfonts}
\usepackage{amsmath}
\usepackage{amssymb}
\usepackage{array}
\usepackage{verbatim}
\usepackage{bm}
\usepackage{epsfig}
\usepackage{graphicx,color}
\usepackage{relsize}
\usepackage{lineno}
\usepackage{float}
\usepackage{multirow}
\RequirePackage{xspace}

\def\pom{{I\!\!P}}

\newcommand{\rr}{\mbox{\boldmath $r$}}
\newcommand{\rd}{\mbox{\boldmath $\Delta$}}

\newcommand{\rb}{\mbox{\boldmath $b$}}

\begin{document}


\title{\boldmath  Probing the spatial  distribution of gluons within the proton \\ in the coherent vector meson production at large - $|t|$}

\author{Victor P. Gon\c{c}alves}
\email{barros@ufpel.edu.br}
\affiliation{Institute of Physics and Mathematics, Federal University of Pelotas, \\
  Postal Code 354,  96010-900, Pelotas, RS, Brazil}

\author{Bruno D. { Moreira}}
\email{bduartesm@gmail.com}
\affiliation{Departamento de F\'isica, Universidade do Estado de Santa Catarina, 89219-710 Joinville, SC, Brazil.}

\author{Luana  {Santana}}
\email{luanas1899@gmail.com}
\affiliation{Departamento de F\'isica, Universidade do Estado de Santa Catarina, 89219-710 Joinville, SC, Brazil.}
\affiliation{Institute of Physics and Mathematics, Federal University of Pelotas, \\
  Postal Code 354,  96010-900, Pelotas, RS, Brazil}

\begin{abstract} 
The coherent production of vector mesons in photon - hadron interactions is considered one of the most promising observables to probe the  QCD dynamics at high energies and the transverse spatial distribution of the gluons in the hadron wave function.  In this paper, we  perform an exploratory study about the dependence of the transverse momentum  distributions, $d\sigma/dt$,  on the model assumed for the proton density profile. We consider a set of non-Gaussian profiles and include them in the forward dipole - proton scattering amplitude associated with the IP-sat model.  We  demonstrate that  the predictions for $d\sigma/dt$ are similar in the HERA kinematical range, but are very distinct for values of $t$ that will be probed in future colliders. In particular, the presence of diffractive dips in the $t$-distributions associated with the production of $J/\Psi$ and $\rho$ mesons is strongly dependent on the model assumed for the density profile. Similar conclusions are also derived when the non-linear effects on the dipole - target interaction are disregarded. Our results indicate that a future study of the coherent vector meson production at large - $t$ will be useful to constrain the spatial distribution of gluons within the proton.
\end{abstract}

\maketitle

\section{Introduction}
  
\begin{figure}
	\centering
	\includegraphics[width=4.5in]{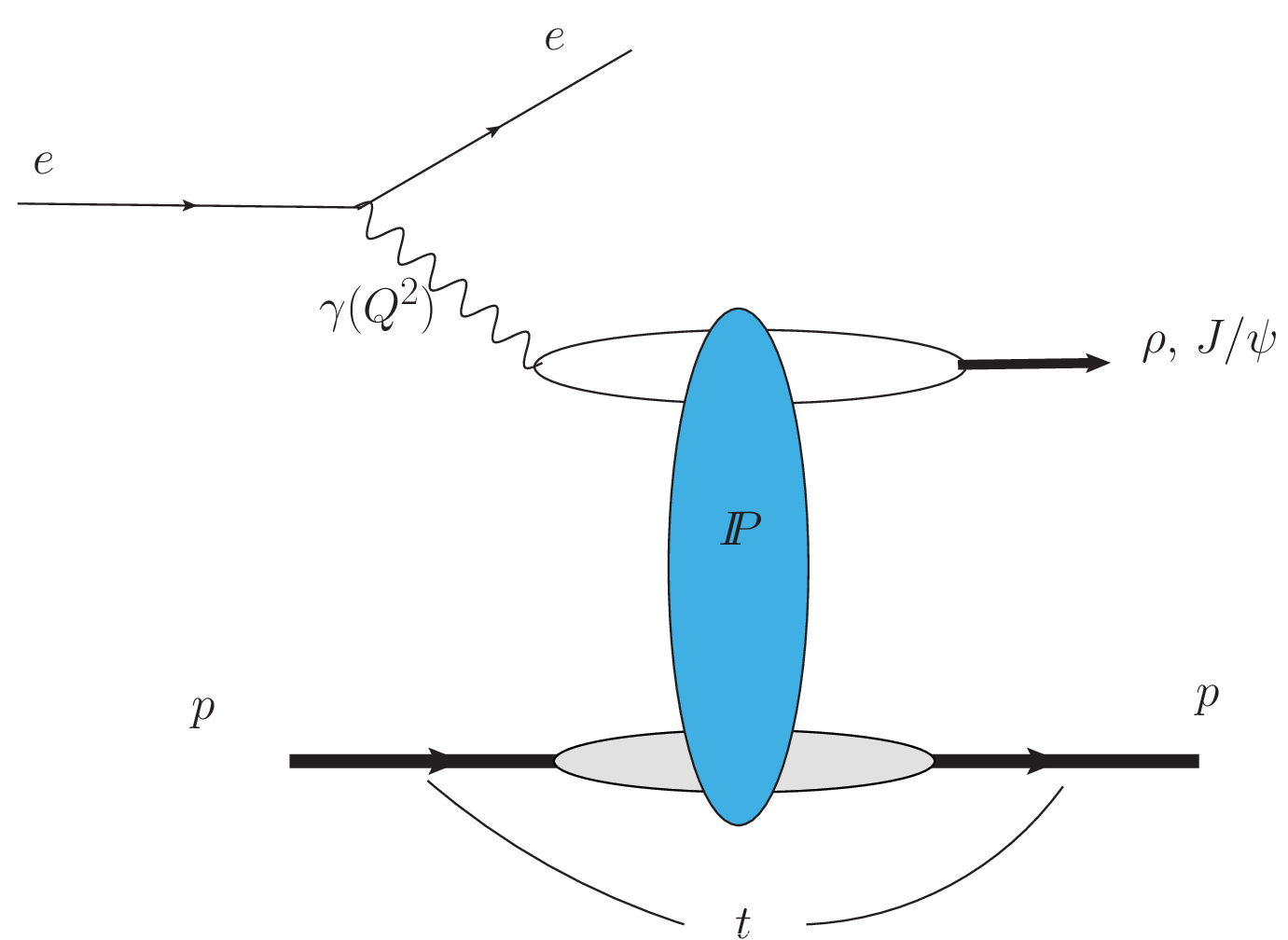}
	\caption{Coherent vector meson production in electron - proton collisions.		 }
	\label{Fig:diagram}
\end{figure}

Over the last decades,   the vector meson production in photon - Pomeron ($\gamma \pom$) interactions has been largely studied, mainly motivated by the possibility to constrain the description of the QCD dynamics at high energies and  improve our understanding of the quantum 3D imaging of the partons inside the protons and nuclei \cite{Klein:2019qfb,Mantysaari:2020axf}. Such a process is  characterized by the presence of a rapidity gap in the final state, due to the color singlet exchange and can be classified as being   coherent or incoherent, depending on if the hadron remains intact or scatters inelastically, respectively. In this paper, we will focus on the coherent vector meson production in electron - proton ($ep$) collisions, represented in Fig. \ref{Fig:diagram}. In this process, the proton scatters elastically and 
remains intact in the final state and  the vector meson is mainly produced with a small squared transverse momentum
$|t|$,  with its   signature being a sharp forward diffraction peak in the transverse - momentum distribution, which has a typical $e^{-B_p.|t|}$ behavior at small - $|t|$, where the quantity $B_p$ can be related with the transverse proton radius. In addition, such a process is expected to provide information on the transverse spatial distribution of
gluons in the proton \cite{Frankfurt:2010ea}. The current data have already  constrained several aspects of the theoretical description of this process, and 	 more precise measurements are expected in the forthcoming years, especially at larger transverse momentum. { Such an expectation motivates the investigation of  how sensitive are the predictions for the $t$- distributions associated with the coherent vector meson production at the future $ep$ colliders to the assumption for the proton density profile. In particular, in this paper, we will consider non - Gaussian profiles and present predictions for the $t$ - distributions considering the kinematical range that will be probed by the  EIC \cite{eic} and LHeC \cite{lhec}.}

{ This paper is organized as follows. In the next Section, we will present a brief review of the color dipole formalism for the exclusive vecton meson production in $ep$ collisions. Moreover, the non - Gaussian profiles and   main assumptions considered in our analysis will be presented. 
In Section \ref{sec:res}, we will compare our results with the HERA data for the electroproduction of $J/\Psi$ and $\rho$ mesons, and present our predictions for the $t$- distributions considering the EIC and LHeC energies. Results will be presented for two different models for the dipole - proton scattering amplitude. Finally, in Section \ref{sec:sum}, we will summarize our main results and conclusions.}
 
\section{Formalism}

The coherent vector meson production in $ep$  collisions is usually described in the color dipole formalism, which  predicts that  the scattering amplitude  can be factorized in terms of the fluctuation of the  virtual photon into a $q \bar{q}$ color dipole, the dipole-proton scattering by a color singlet exchange ($\pom$)  and the recombination into the exclusive final state, being given by (See, e.g. Ref. \cite{KMW})
\begin{eqnarray}
 {\cal A}_{T,L}({x},Q^2,\Delta)  =  i
\,\int d^2\rr \int \frac{dz}{4\pi} \int \, d^2\rb \, e^{-i[\rb -(1-2z)\rr/2].\rd}
 \,\, (\Psi^{V*}\Psi)_{T,L}  \,\,\frac{d\sigma_{dp}}{d^2\rb}({x},\rr,\rb)
\label{amp}
\end{eqnarray}
where $T$ and $L$ denotes the transverse and longitudinal polarizations of the virtual photon, $\Delta = \sqrt{-t}$ is the momentum transfer, $Q^2$ is the photon virtuality and $x = (M^2 + Q^2 -t)/(W^2+Q^2)$, with  $W$ being the center of mass energy of the virtual photon -- proton system and $M$  the mass of the vector meson. 
The variables  $\rr$ and $z$ are the dipole transverse radius and the momentum fraction of the photon carried by a quark (an antiquark carries then $1-z$), respectively, and   $\rb$ is the impact parameter of the dipole relative to the proton. Moreover,    $(\Psi^{V*}\Psi)_i$ denotes the wave function overlap between the virtual photon and the vector meson wave functions and ${d\sigma_{dp}}/{d^2\rb}$  is the dipole-proton cross-section (for a dipole at  impact parameter $\rb$) which encodes all the information about the hadronic scattering, and thus about the non-linear and quantum effects in the hadron wave function \cite{hdqcd}. The differential cross-section for coherent  interactions is given by
\begin{equation}\label{eq:xsec-coh}
  \left.\frac{d\sigma^{\gamma^* p \rightarrow V \,p}}{dt}\right|^{coh}_{T,L} =
  \frac{1}{16\pi}\left| \left\langle \mathcal{A}_{T,L}(x,Q^2, \Delta) \right\rangle \right|^2\,\,,
\end{equation}
where $\left\langle ... \right\rangle$  represents the average over the configurations of the proton wave function. As the average is performed at the  level of the scattering amplitude, one has that coherent interactions probe the averaged density profile of the gluon density \cite{Toll,Armesto:2014sma,Mantysaari:2016ykx,Mantysaari:2016jaz,cepila}.

The $t$ - distribution for the coherent vector meson production is determined by the dependence of ${d\sigma_{dp}}/{d^2\rb}$ on the impact parameter, which is associated with the QCD dynamics and/or the spatial distribution of gluon within the proton. The description of this dependence is a theme of intense debate, since the first studies of the Balitsky - Kovchegov (BK) equation taking into account of the impact parameter dependence of the forward dipole scattering amplitude (${\cal{N}}({x},\rr,\rb) = \frac{1}{2} {d\sigma_{dp}}/{d^2\rb}$) performed in Refs. \cite{Golec-Biernat:2003naj,Berger:2010sh,Berger:2011ew,Berger:2012wx,Mantysaari:2018zdd}   have  found out that the solutions acquired a so-called Coulomb tail, meaning that
the contribution at large impact parameters grew too fast and a non-perturbative contribution must be included in order to describe the HERA data for the $F_2$ structure function.  More recently, in Refs. \cite{Cepila:2018faq,Bendova:2019psy}, the BK equation was solved for a proton target including the dependence on impact parameter and using the collinearly-improved kernel. These studies have demonstrated that the contribution coming from the large impact parameters is strongly suppressed by the collinear corrections and that the experimental data for  inclusive, diffractive and exclusive observables at HERA are reasonably well described \cite{Cepila:2018faq,Bendova:2019psy,Bendova:2020hkp,Bendova:2022xhw}. It is important to emphasize that this solution has been obtained assuming an initial condition characterized by a Gaussian distribution for the impact parameter dependence of the matter within the proton.
Another alternative to describe the HERA data for exclusive processes is to assume a phenomenological model for  ${\cal{N}}(x,\rr,\rb)$ based on approximations for the Color Glass Condensate (CGC) formalism.
Two examples of very successful models are the b-CGC \cite{Watt_bCGC,KMW} and IP-sat \cite{ipsat1} models, which  have been updated in Refs. \cite{Rezaeian_update,ipsat4,ipsat_heikke,Sambasivam:2019gdd} using the high precision HERA data to constrain their 
free parameters and describe the data quite well. 
In particular,   IP-sat model  incorporates the saturation effects via the Glauber - Mueller approximation \cite{mueller}, assuming an eikonalized 
form for ${\cal N}$  that depends on a gluon distribution evolved via DGLAP equation and that resums higher twist contributions. The associated  dipole - proton scattering amplitude is  given 
by \cite{ipsat_heikke}
\begin{eqnarray}
 {\cal N}(x,\mbox{\textbf{\textit{r}}},\mbox{\textbf{\textit{b}}}) = 
 1 - \exp \left[-
\frac{\pi^{2}r^{2}}{2 N_{c}} \alpha_{s}(\mu^{2}) \,\,xg\left(x, \mu_{0}^{2} + \frac{C}{r^{2}}  
\right)\,\, T_p(\rb) 
 \right] ,
 \label{ipsat}
\end{eqnarray}
and  a Gaussian profile
\begin{eqnarray}
T_p(\rb) = \frac{1}{2\pi B_{p}}  
\exp\left(-\frac{\rb^{2}}{2B_{p}} \right) 
\end{eqnarray}
is assumed for the proton density profile.
In Ref. \cite{ipsat_heikke}, { the authors have assumed  $B_p = 4$ GeV$^{-2}$ 
based on a fit of the HERA exclusive $J/\Psi$ production data, since the inclusive data is not very sensitive to $B_p$ \footnote{In Ref. \cite{ipsat_heikke}  the authors have neglected the off-forward correction to the vector meson wave function, which is included in our analysis. As a consequence, a slightly larger value is expected if a new fit of the data is performed.}. It is important to emphasize that the
effective transverse area of the proton is not a free parameter
in the  model, since  the root-mean-square radius of the proton is given by $\sqrt{2B_p}$.}
The initial gluon distribution evaluated at $\mu_{0}^{2}$ is taken to be 
\begin{eqnarray}
xg(x,\mu_{0}^{2}) =  A_{g}x^{-\lambda_{g}} (1-x)^{6} .
\end{eqnarray}
The free parameters of this model are fixed by a fit of HERA data.  The calculations will be performed assuming that the overlap functions $(\Psi^{V*}\Psi)_i$ are described by the Gaus-LC model (For a detailed discussion see, e.g., Refs. \cite{Goncalves:2004bp,Goncalves:2017wgg}).

A Gaussian profile for $T_p(\rb)$ is a reasonable ansatz in the kinematical range probed by HERA, since all diffractive vector-meson $t$ distributions {  can be approximately described  as follows
\begin{eqnarray}
 \left.\frac{d\sigma^{\gamma^* p \rightarrow V \,p}}{dt}\right|^{coh} \propto \exp(-B_p \cdot |t|) \,\,,
\end{eqnarray}
{ with the exponent $B_p$ being dependent on the vector meson}, photon virtuality $Q^2$ and photon - hadron center - of - mass energy $W$ (See, e.g., Refs.  \cite{ZEUS:2004yeh,ZEUS:2007iet}).}
However, the HERA data is limited to small values of $t$ ($\lesssim 1.0$ GeV$^2$). In contrast, the future electron - hadron colliders at the BNL and CERN \cite{eic,lhec} are expected to probe larger values of $t$ and provide a larger amount of data in this kinematical region due to the high projected luminosities. Considering this perspective, a natural question arises: How sensitive are the predictions for the $t$- distributions associated with the coherent vector meson production at the EIC and LHeC to the assumption for the density profile $T_p(\rb)$?
In other words, can the future data be used to constrain the spatial distribution of gluons within the proton?
Our goal in this exploratory study is to obtain a first answer for this question by analyzing how a non - Gaussian profile in the IP-sat model modifies its predictions for the coherent $J/\Psi$ and $\rho$ electroproduction at large - $t$. Motivated by Refs. \cite{Rybczynski:2013mla,Lappi:2023frf}, we will parametrize the  density profile using the
regularized incomplete gamma function profile as follows:
\begin{eqnarray}
    T_p(\rb)=\frac{1}{2\pi B_p}\frac{\Gamma\left(\frac{1}{\omega},\frac{\rb^2}{2B_p\omega}\right)}{\Gamma(1/\omega)} \,\,,
    \label{tb}
\end{eqnarray}
where  the parameter $\omega$ controls the steepness of the  profile.
Such a parametrization  has as limits the hard - sphere profile for $\omega \rightarrow 0$ and the Gaussian one for $\omega \rightarrow 1$, and allow us to explore other non - Gaussian profiles.  { As in Refs. \cite{Armesto:2014sma,Watt_bCGC,ipsat1, Rezaeian_update,ipsat4,ipsat_heikke}, we will disregard that $B_p$ can be dependent on $Q^2$ and $W$ (See, e.g., Refs. \cite{Berger:2012wx,Mantysaari:2018zdd} for a discussion about the energy dependence of the slope).} 
In our analysis, we will modify the FORTRAN library provided by the authors of Ref. \cite{ipsat4} in order to  
calculate ${\cal N}$ for the density profile given by Eq. (\ref{tb}), considering different values of $\omega$. The value of $B_p$ for each $\omega$ will be constrained by the $J/\Psi$ HERA data, and we will present the predictions for the $t$-distributions considering the kinematical range that will be probed by the EIC and LHeC. { It is important to emphasize that the maximum value of $t$ that will be probed by these colliders is still an open question, since it is strongly dependent on the detector characteristics. In general, it is expected that they will be able to measure coherent events with values of $t$ larger than those probed at HERA, with a conservative expectation being that these detectors will cover the kinematical range of $|t| \le 2$ GeV$^2$. However, it is usual in phenomenological studies to present the predictions for a  larger $|t|$ range in order to demonstrate the presence (or not) of secondary dips and motivate future measurements of coherent events in this kinematical range (see, e.g. Ref. \cite{Mantysaari:2020lhf}). In our analysis, we will present the predictions for $d\sigma/dt$ in the range $|t| \le 7$ GeV$^2$, since it allow us to demonstrate the impact of the different profiles functions in the position of the dips, when they are present, but it does not means that the future $ep$ colliders will be able to cover this kinematical range. As we will demonstrate in the next section, the modeling of the profile functions already affects the behavior of the distribution for $|t| \approx 2$ GeV$^2$. As a consequence, the future colliders should be, in principle, able to constrain the description of the proton density profile. }

\begin{figure}
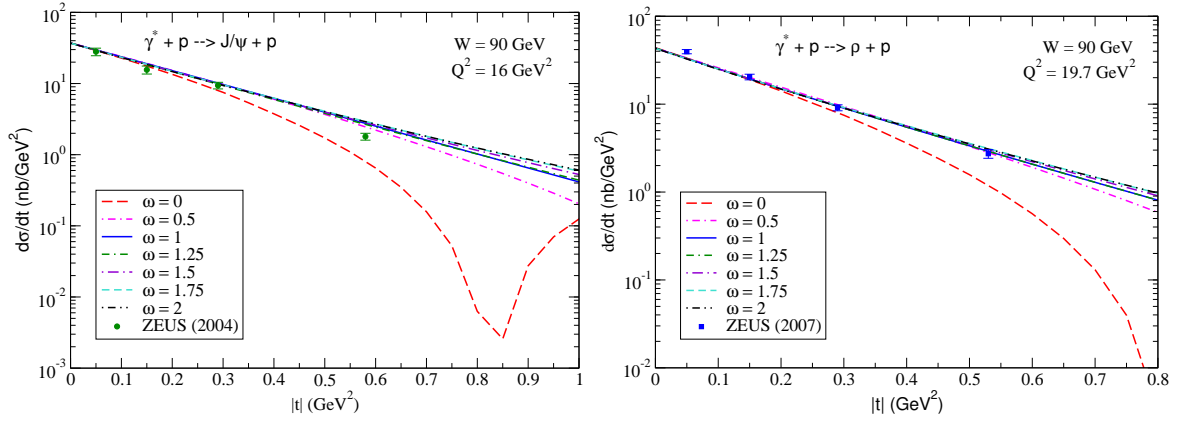

	\centering
	\includegraphics[width=3in]{jpsi_W90_Q216.eps}
	\includegraphics[width=3in]{rho_W90_Q219_7.eps}
		\caption{{\bf Left panel:} Predictions of the IP-sat model for  $d\sigma/dt$ considering the electroproduction of $J/\Psi$ mesons, derived assuming distinct values of $\omega$. Data from Ref.  \cite{ZEUS:2004yeh}.   {\bf Right panel:} Comparison between the predictions  for the electroproduction of $\rho$ mesons and the ZEUS data \cite{ZEUS:2007iet}. }
	\label{Fig:HERAdata}
\end{figure}

{ Before to present our results, some comments are in order. In our analysis, we are only considering the coherent production of vector mesons. In particular, we assume that the events associated with the coherent processes can be separated by tagging the protons in the final state using very forward detectors, as currently made in the study of exclusive processes at the LHC.  However, at large values of $t$ the dominant process is the incoherent scattering, in which the proton gets excited and subsequently dissociates. In recent years, several groups \cite{Mantysaari:2016ykx,Mantysaari:2016jaz,Traini:2018hxd,Cepila:2017nef,Cepila:2016uku,Cepila:2018zky,Kumar:2022aly,Kumar:2021zbn,Demirci:2022wuy,Blaizot:2022bgd,Cepila:2023dxn,Xiang:2023msj,Kumar:2024kns}  have demonstrated that the description of the incoherent production is sensitive to geometrical fluctuations in the proton’s transverse gluonic wave-function and have considered different approaches to treat the number and size of gluon density hotspots within the proton. Such approaches have derived very interesting predictions, which are expected to be probed in the forthcoming years, which will be important to constrain the proton structure. In this exploratory study, we will not take into account of this geometrical fluctuations, since we are only interested in investigating how sensitive are the predictions for the coherent production to the model assumed for the proton profile. Surely, the description of both coherent and incoherent processes should be considered in a unified way in order to improve our understanding of the spatial distribution of gluons within the proton, but it is beyond the scope of the current study. Another important aspect is that in our analysis, we are only considering models for the proton profile function that are usually considered in the literature. However, a more precise way would be taken different three-dimensional shapes of the proton, and then integrate over the longitudinal direction. We intend to implement this improvement in a future study. Although the predictions could be modified by these improvements, we believe that the main conclusion of this paper, that the analysis of the coherent production at large - $t$ could be useful to constrain the proton structure, will remain valid.   }

\section{Results}
\label{sec:res}

Initially, in Fig. \ref{Fig:HERAdata} (left panel) we present the predictions of the IP-sat model for the $t$ - distribution associated with the electroproduction of $J/\Psi$ mesons for $W = 90$ GeV and $Q^2 = 16$ GeV$^2$.  The results for different values of $\omega$ are shown and compared with the ZEUS data \cite{ZEUS:2004yeh}. As  we are interested in the dependence on the squared transverse momentum of the distribution, we have fixed the value of $B_p$ for distinct $\omega$ such that all predictions coincide for $|t| = 0$. In a forthcoming study we plan to perform a fit of the HERA data for { exclusive} observables considering $B_p$ and $\omega$ as fitting parameters. However, in this exploratory study, we believe that a naive approach is enough to verify the dependence of the predictions on the density profile assumed in the calculations. The associated values of $B_p$ for the IP-sat model are presented in Table \ref{tab:bp}. The results presented in Fig. 
\ref{Fig:HERAdata} (left panel) indicate that the hard sphere model ($\omega = 0$) implies a much more steeply falling than the other models with the increasing of $t$, predicting a dip at $|t| \approx 0.85$ GeV$^2$. Such a behavior already had been derived in Ref. \cite{ipsat1} and this simplified model for the density profile was discarded since it fails to describe the HERA data for exclusive observables.  In contrast, the predictions derived assuming the other values of $\omega$ are similar in the kinematical range probed by HERA. { Similar conclusions are also obtained when we compare our predictions  with the experimental $J/\Psi$ data for $Q^2 = 0$ (not shown).}  As a cross-check, in Fig. \ref{Fig:HERAdata} (right panel), we compare our predictions, derived assuming the same values of $B_p$ fixed by the $J/\Psi$ data, with the ZEUS data for the electroproduction of $\rho$ mesons for $W = 90$ GeV and $Q^2 = 19.7$ GeV$^2$ \cite{ZEUS:2007iet}. One has that the data is quite well described if $\omega \neq 0$ is assumed.  Finally, it is important to emphasize that one has verified that the predictions for inclusive observables at HERA as e.g.  for the proton structure function,  are very similar if we assume a non - Gaussian profile and the values of $B_p$ and $\omega$ presented in Table \ref{tab:bp}, { with the normalization being modified by less than 3\% in the small - $x$ HERA kinematics.}

\begin{table}[t]
    \centering
    \begin{tabular}{|c|c|c|c|c|c|c|c|} \hline
  $B_p$ (GeV$^{-2}$) &  $\omega = 0$ & $\omega = 0.5$ & $\omega = 1.0$ &
$\omega = 1.25$ & $\omega= 1.5$ & $\omega= 1.75$ & $\omega = 2.0$ \\  
\hline
IP-sat & 8.8   &  5.2   & 4.0    & 3.7  & 3.3  & 3.0    & 2.8   \\   
\hline
IP-nonsat & 8.4   &  5.0   & 4.0    & 3.7  & 3.4  & 3.1   &  2.9  \\
\hline
    \end{tabular}
\caption{Values of  $B_p$ (GeV$^{-2}$) constrained by the $J/\Psi$ HERA data using the IP-sat and IP-nonsat models. }
\label{tab:bp}
\end{table}

\begin{figure}
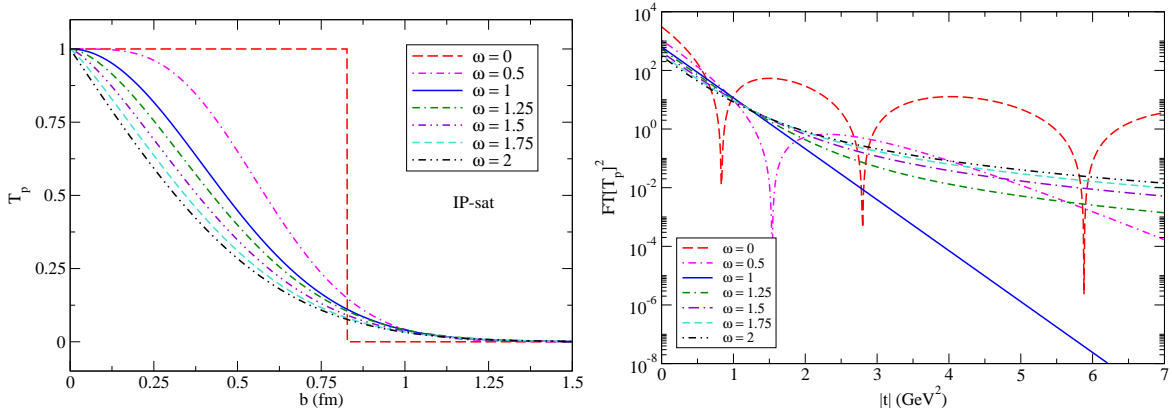

	\centering
	\includegraphics[width=3in]{perfil_proton_ipsat.eps}
	\includegraphics[width=3in]{transformada.eps}
		\caption{{\bf Left panel:} Impact parameter dependence of the density profile $T_p(\rb)$ for different values of $\omega$.  {\bf Right panel:} Predictions for the square of the Fourier transform of $T_p(\rb)$. }
	\label{Fig:profile}
\end{figure}

In Fig. \ref{Fig:profile} (left panel) we present the impact parameter dependence of the density profile $T_p(\rb)$ for different values of $\omega$ considering the values of $B_p$ for the IP-sat model. As expected for a hard sphere model ($\omega = 0$), the associated profile has a step function behavior, $\Theta(R_p - b)$, with $R_p = 0.828$ fm. On the other hand, for $\omega = 0.5$, one has a profile that predicts a slower decreasing with $b$ than the Gaussian ($\omega = 1$) close to the center of the proton. Such a behavior is present for all profiles characterized by $\omega < 1$ (not shown). In contrast, for $\omega > 1$, the decreasing with the impact parameter is faster than the Gaussian profile for small $b$. The  square of the Fourier transform of the density profile is presented in Fig. \ref{Fig:profile} (right panel). One has that dips in the distribution are predicted for $\omega < 1.0$. In contrast, for $\omega \ge 1.0$, we do not obtain any dips, in agreement with the results derived in Ref. \cite{Lappi:2023frf}, with the predictions for large values of $|t|$ being distinct.

\begin{figure}
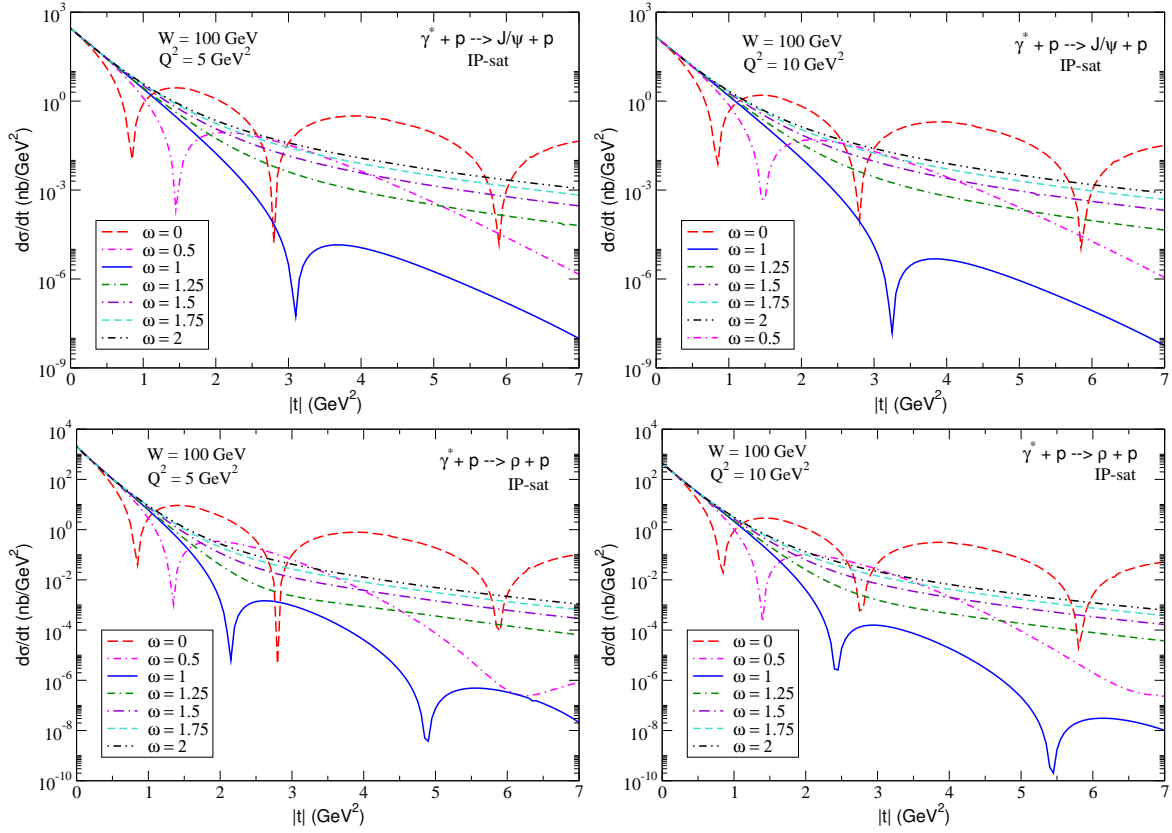

	\centering
	\includegraphics[width=3in]{jpsi_ipsat_W100_Q2_5.eps}
	\includegraphics[width=3in]{jpsi_ipsat_W100_Q2_10.eps}
		\includegraphics[width=3in]{rho_ipsat_W100_Q2_5.eps}
		\includegraphics[width=3in]{rho_ipsat_W100_Q2_10.eps}

		\caption{Predictions of the IP-sat model for the electroproduction of $J/\Psi$ (upper panels) and $\rho$ (lower panels) mesons considering $W = 100$ GeV and  $Q^2 = 5.0$ GeV$^2$ (left panels) and 10.0 GeV$^2$ (right panels).}
	\label{Fig:dsdtw100}
\end{figure}

In Fig. \ref{Fig:dsdtw100}  we present our predictions for the transverse momentum distribution for the electroproduction of $J/\Psi$ (upper panels) and $\rho$ (lower panels) mesons,  derived using the IP-sat model and considering different values of $\omega$. The results were derived for $W = 100$ GeV and  $Q^2 = 5.0$ GeV$^2$ (left panels) and 10.0 GeV$^2$ (right panels). For a Gaussian profile ($\omega = 1$), the IP-sat model predicts the presence of diffractive dips in the $t$ - distribution in the range considered, with the number of dips being larger for the $\rho$ meson. It is important to emphasize that for this value of $\omega$, dips are not present in the Fourier transform of $T_p(\rb)$. For $\omega < 1.0$, dips are present in both the Fourier transform and the $t$ - distribution. In contrast, for $\omega > 1.0$, 
dips are not predicted, with the normalization of the distribution  for large - $t$ being dependent on the density profile assumed. { One has verified that similar results are obtained for the LHeC energy ($W = 1000$ GeV)}. Such results indicate that future experimental data from EIC and LHeC, { in the range $|t| \approx 2$ GeV$^2$}, will be able to constrain the modeling of the density profile and, therefore, to probe the spatial distribution of gluons within the proton.

\begin{figure}
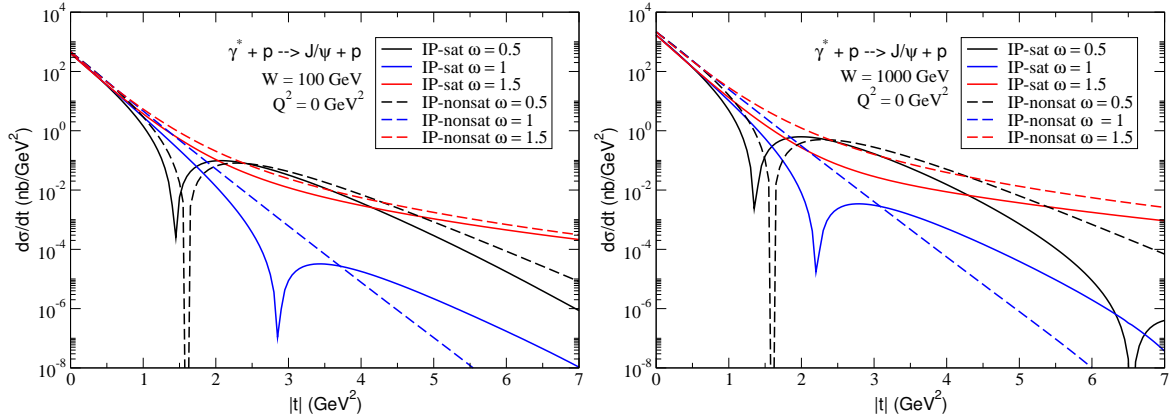

	\centering
	\includegraphics[width=3in]{jpsi_W100_Q2_0.eps}
	\includegraphics[width=3in]{jpsi_W1000_Q2_0.eps}
		\caption{Comparison between the predictions of the IP-sat and IP-nonsat models for the photoproduction of $J/\Psi$ mesons at $W = 100$ GeV (left panel) and $W = 1000$ GeV (right panel).}
	\label{Fig:comparison}
\end{figure}

Before to summarize our main results, it is important to analyze the dependence of our predictions on the model assumed for the dipole - proton scattering amplitude. The previous results were derived considering the IP-sat model, which takes into account of non-linear effects in the QCD dynamics. We have repeated our analysis considering a linearized  version of this model, usually denoted IP-nonsat model, which disregard these non-linear effects and has a dipole - proton scattering amplitude given by \cite{ipsat_heikke}
\begin{eqnarray}
 {\cal N}(x,\mbox{\textbf{\textit{r}}},\mbox{\textbf{\textit{b}}}) = 
\frac{\pi^{2}r^{2}}{2N_{c}} \alpha_{s}(\mu^{2}) \,\,xg\left(x, \mu_{0}^{2} + \frac{C}{r^{2}}  
\right)\,\, T_p(\rb) \, \,.
 \label{ipnonsat}
\end{eqnarray}
We have considered the parameters obtained in Ref. \cite{ipsat_heikke} for this dipole model, and fixed the values of $B_p$ for the distinct $\omega$ using the $J/\Psi$ HERA data. The corresponding values are presented in Table \ref{tab:bp}. For this model, one has that the $t$ - distribution for the vector meson production will be fully determined by the square of the Fourier transform of $T_p(\rb)$, in contrast with the IP-sat model, where the density profile is in the exponential [See Eq. (\ref{ipsat})]. The comparison between the predictions of the IP-sat and IP-nonsat models for the photoproduction of $J/\Psi$ mesons is  presented in Fig.  \ref{Fig:comparison}, where we show the results for  
$W = 100$ GeV (left panel) and $W = 1000$ GeV (right panel). While for a Gaussian profile, one has a clear difference between the IP-sat and IP-nonsat predictions, for the other values of $\omega$ the distributions predicted by these two models are similar. In particular, one has that for $\omega = 0.5$, both models predict a dip in the distribution for $|t| \approx 1.5$ GeV$^2$. In contrast, for $\omega = 1.5$, a dip is not predicted by both models, with the results  being different only in the normalization.

\section{Summary}
\label{sec:sum}

The description of the proton structure  is still one of the main challenges  of Particle Physics. In last decades, our understanding about the subject has largely improved, mainly associated with the large amount of data released by HERA and LHC. One of the more promising observables is the exclusive production of vector mesons, which  is strongly sensitive to the underlying QCD dynamics, since this process is driven by the gluon content of the target, and to the transverse spatial distribution of the gluons in the hadron wave function. The investigation of these aspects is one of the main motivations for the future electron - hadron colliders, which will probe unexplored kinematical region and are characterized by higher luminosities. In this paper, we have  performed an exploratory study about the dependence of the $t$ - distributions on the model assumed for the proton density profile. In particular, we have considered a set of non-Gaussian profiles and included them in the forward dipole - proton scattering amplitude of IP-sat model. The free parameter present in these new profiles was fixed using the $J/\Psi$ HERA data. We have demonstrated that the predictions for $d\sigma/dt$  from different models are similar in the $t$ - range studied by HERA, but differ for values of $t$ that will be probed in the future colliders. Moreover, the presence of diffractive dips in the $t$-distributions associated with the production of $J/\Psi$ and $\rho$ mesons is strongly dependent on the model assumed for the density profile. Similar conclusions are also derived when the non-linear effects on the dipole - target interaction are disregarded. All these results indicate that a future study of the coherent vector meson production at large - $t$ will be useful to constrain the spatial distribution of gluons within the proton.

\section*{Acknowledgments}
 V.P.G. was partially supported by  CNPq,  FAPERGS and  INCT-FNA (process number 
464898/2014-5). B. D. Moreira was partially supported by FAPESC.  L. Santana was partially supported by FAPESC, PROMOP/UDESC and CAPES (Finance Code 001).

\end{document}